\documentclass{ws-p8-50x6-00}
\newcommand{\nn}{\nonumber\\ }
\newcommand{\beq}{\begin{eqnarray}}
\newcommand{\eeq}{\end{eqnarray}}
\def\labe{\label}
\def\simge{\mathrel{%
   \rlap{\raise 0.511ex \hbox{$>$}}{\lower 0.511ex \hbox{$\sim$}}}}
\def\simle{\mathrel{
   \rlap{\raise 0.511ex \hbox{$<$}}{\lower 0.511ex \hbox{$\sim$}}}}
\def\bigs{\mathrel{
   \rlap{\raise 0.531ex \hbox{$>$}}{\lower 0.531ex \hbox{$<$}}}}

\def\del{\partial}                              

\begin{document}

\title{Gluon Saturation at small x}

\author{Edmond Iancu}

\address{Service de Physique Th\'eorique, CE Saclay,
        F-91191 Gif-sur-Yvette, France\\
E-mail: eiancu@cea.fr}


\maketitle

\abstracts{At very high energies, the relevant component of the
hadron wavefunction can be described as a Color Glass Condensate, 
i.e., a state of high density gluonic matter whose distribution 
is random, but frozen over the relevant time scales. The weight
function for this distribution obeys a renormalization 
group equation in the form of a functional Fokker-Planck equation.
Its solution leads to an effective theory which predicts gluon saturation 
at sufficiently high energy, or small Bjorken's $x$.
}

\section{Introduction}

Hadronic scattering at high energy, or small Bjorken's $x$, uncovers
a novel regime of QCD where the coupling is small ($\alpha_s\ll 1$) but 
the parton densities are so large that conventional perturbation theory 
breaks down, via strong non-linear effects 
\cite{GLR,AM0,MV94} (see \cite{Larry01,Levin01} for recent reviews and
more references). Remarkable progress in recent years has resulted
from the observation that this high density
regime can be studied via semiclassical methods\cite{MV94}. 
In what follows, I shall describe a classical
effective theory\cite{MV94,JKLW97,PI,SAT} 
 which is well suited to study the non-linear phenomena
at small $x$ and has a transparent physical interpretation: It portrays 
the gluon component of the hadron wavefunction (the relevant
component at small $x$) as a {\it Color Glass Condensate} (CGC).
This is a multiparticle quantum state with 
high occupation numbers, but to the accuracy of interest it can be 
represented as a stochastic color field 
with a probability law determined by a functional Fokker-Planck equation.

The latter is a renormalization group equation\cite{JKLW97,PI}
(RGE) which shows how to
construct the effective theory by integrating out quantum fluctuations
in the background of a strong color field (the CGC).
The non-linear effects included in the RGE via this background field
describe interactions among the gluons 
produced in the quantum evolution towards small $x$. 
This leads to a non-linear generalization of the BFKL equation 
which, remarkably, predicts\cite{SAT} the {\it saturation} of 
the gluon distribution at small $x\,$: Unlike the BFKL result
for the gluon distribution, which increases exponentially with the
rapidity $\tau\equiv \ln(1/x)\propto \ln s$ and thus violates
the Froissart unitarity bound $\sigma\le \ln^2 s$, 
our corresponding prediction grows only linearly with $\tau$,
thus being consistent with the unitarity bounds.

\section{The effective theory for the CGC}
\label{sec:EFT}

The effective theory applies to gluon correlations in the hadron 
 wavefunction as measured in deep inelastic 
scattering at small Bjorken's $x$. It is formulated in the hadron
infinite momentum frame, where small $x$ corresponds to soft
longitudinal momenta\footnote{I
use light-cone vector notations, e.g.,
$k^\mu=(k^+,k^-,k_\perp)$, with
$k^+\equiv (k^0+k^3)/\sqrt 2$,
$k^-\equiv (k^0-k^3)/\sqrt 2$, ${k}_\perp\equiv (k^1,k^2)$,
and $k\cdot x = k^- x^+ + k^+ x^-
- { k}_\perp \cdot {x}_\perp$.} 
$k^+=xP^+$, with $P^+$ the hadron momentum, and $x\ll 1$.
The main observation is that the ``fast''
partons (i.e., the excitations with
$p^+\gg k^+$) can be replaced,
as far as their effects on the soft correlation functions are 
concerned, by a classical random {\it color source} $\rho^a(x)$, 
whose gross properties are determined by the soft--fast
separation of scales:

The fast partons
appear to the soft gluons as a color charge distribution $\rho^a(x)$ 
which is {\it static} (i.e., independent of $x^+$), {\it localized} near 
the light-cone (within a small distance $\Delta x^- \sim 1/p^+ \ll 1/k^+$), 
and {\it random} (since this is the instantaneous color charge in the
hadron ``seen'' by the soft gluons at the arbitrary time of their
emission), with gauge-invariant probability density $W_\tau[\rho]$.
(We use the rapidity 
$\tau\equiv\ln(P^+/k^+) = \ln(1/x)$ to indicate the
dependence of the weight function upon the soft scale $k^+$.)

Gluon correlations at the scale $k^+ =
xP^+$ are obtained as (${\vec x}\equiv (x^-,{x}_{\perp})$):
\beq\label{eq:clascorr}
\langle A^i_a(x^+,\vec x)A^j_b(x^+,\vec y)
\cdots\rangle_\tau\,=\,
\int {\cal D}\rho\,\,W_\tau[\rho]\,{\cal A}_a^i({\vec x})
{\cal A}_b^j({\vec y})\cdots\,,\eeq
where ${\cal A}_a^i\equiv {\cal A}_a^i[\rho]$ is the 
solution to the classical Yang-Mills equations 
\beq
(D_{\nu} F^{\nu \mu})_a(x)\, =\, \delta^{\mu +} \rho_a(x)\,
\label{cleq0}
\eeq
in the light-cone 
gauge $A^+_a=0$, which is the gauge which allows for the most
direct contact with the gauge-invariant physical quantities 
\cite{AM0,PI}. For instance, the gluon distribution function
is obtained as 
\beq\label{GCL}
x G(x,Q^2)&=&\frac{1}{\pi}
\int {d^2k_\perp \over (2 \pi)^2}\,\Theta(Q^2-
k_\perp^2)\,\Bigl\langle\,
|{\cal F}^{+i}_a(\vec k)|^2\Bigr\rangle_\tau\,.\eeq
where ${\cal F}^{+i}_a=\partial^+ {\cal A}^i_a$ is the 
electric field associated to the classical solution
${\cal A}_a^i[\rho]$, and $\vec k \equiv (k^+,{k}_\perp)$
with $k^+=xP^+=P^+{\rm e}^{-\tau}$.

Eqs.~(\ref{eq:clascorr}) and (\ref{cleq0}) are those for a {\it glass}
(here, a {\it color} glass): There is a stochastic color charge,
that is averaged over. This is frozen over short time 
scales, of the order of the lifetime 
$\Delta x^+ \sim 1/k^- \propto k^+$ of the soft gluons,
but it changes randomly over the larger time scale
$1/p^- \gg  1/k^-$, which is the natural time
scale for the dynamics of the fast partons. This is like
a glass which is disordered and is a liquid on long time
scales but seems to be a solid on short time scales.
We shall see later that, at saturation, the gluons are very
densely packed, with a phase-space density $\sim 1/\alpha_s$, 
which is the maximum density allowed by their mutual
interactions. They form a {\it Bose condensate}. 

In the effective theory, this {\it Color Glass Condensate}
is described by strong classical color fields
${\cal A}^i_a\sim 1/g$. The classical dynamics
is then fully non-linear, so one needs
the exact solution to eq.~(\ref{cleq0}), which reads\cite{PI}
\beq\label{Aclass}\,
{\cal A}^i\,
(\vec x) &=&{i \over g}\, U(\vec x) \,\partial^i  U^\dagger(\vec x),\\
U^{\dagger}(x^-,x_{\perp})&=&
 {\rm P} \exp
 \Big\{
ig \int_{-\infty}^{x^-} dz^-\,{\alpha}(z^-,x_{\perp})
 \Big \},\label{Udef}\\
- \nabla^2_\perp \alpha({\vec x})&=&\rho(\vec x).
\label{alpharho}\eeq
Since $\rho(\vec x)$, and therefore $\alpha({\vec x})$, are
localized near $x^-=0$, so is the electric field, which appears
effectively as a $\delta$-function to any probe with  low 
longitudinal resolution (i.e., with momenta $q^+ < k^+$) :
\beq\labe{v}{\cal F}^{+i}(\vec x) \approx\delta(x^-)\,
\frac{i}{g}\,V(\del^i V^\dagger),\quad
V^\dagger(x_{\perp})\equiv{\rm P} \exp
 \Big \{
ig \int\!\! dx^-\alpha^a (x^-,x_{\perp})T^a
 \Big \}.\,\,\eeq

\section{The Renormalization Group Equation}
\label{QEVOL}

The weight function $W_\tau[\rho]$ 
for the effective theory at the scale $k^+=xP^+$ 
is obtained by integrating out the quantum fluctuations with
$p^+ > k^+$ in layers of $p^+$, to ``leading logarithmic 
accuracy'' --- i.e., by retaining only the terms 
enhanced by the large logarithm $\ln(1/x)$ ---, but to all orders in the
 background fields ${\cal A}^i_a$ 
generated at the previous steps \cite{JKLW97,PI}.
The background fields are an essential ingredient: their interactions
with the quantum gluons correspond, in the more
conventional picture of the ``parton cascades'' \cite{GLR}, to
rescatterings among the soft gluons radiated in the quantum evolution 
towards small $x$ (e.g., recombinations of gluons from different
cascades).

The quantum evolution can be formulated as a renormalization group
equation (RGE) for the flow of $W_\tau[\rho]$ 
with $\tau=\ln(1/x)$. To motivate the structure
of this equation, I succintly describe one step in this 
RG procedure.
Assume that we know the effective theory at some initial
scale $\Lambda^+$ --- as specified by 
$W_\Lambda[\rho]\equiv W_\tau[\rho]$,
with $\tau=\ln(P^+/\Lambda^+)$ ---, and we are interested
in correlations at the softer scale
$k^+ \sim b\Lambda^+$ with $b\ll 1$ and $\alpha_s\ln(1/b)< 1$.
Our purpose is to construct the new weight function 
$W_{b\Lambda}[\rho]\equiv  W_{\tau+\Delta\tau}[\rho]$
($\Delta\tau\equiv \ln(1/b)$), which would determine the 
gluon correlations at this softer scale. As compared to 
$W_\Lambda[\rho]$, the new weight function must include also the 
quantum effects induced by the ``semi-fast'' gluons 
with $p^+$ in the strip
\beq\labe{strip}\,\,
 b\Lambda^+ \,\,<\,\, |p^+|\,\, <\,\,\Lambda^+\,.\eeq
To characterize these effects, consider the 
change in the 
2-point function $\langle A^i_a(x)  A^i_a(y)\rangle_\Lambda$
with decreasing 
$\Lambda^+$. At the original scale $\Lambda^+$, we can use the effective
theory to write $\langle A^i(x)  A^i(y)\rangle_\Lambda \,=\,
\langle{\cal A}^i(x){\cal A}^i(y)\rangle_{W_\Lambda}$,
with ${\cal A}^i[\rho]$ the classical solution in eq.~(\ref{Aclass}),
and the average over $\rho$ computed as in eq.~(\ref{eq:clascorr}).
At the new scale $b\Lambda^+$, 
$A^i_a={\cal A}^i_a + \delta A^i_a$, with
$\delta A^i_a(x)$ representing the quantum fluctuations with 
momenta $p^+<\Lambda^+$. Thus,
\beq\label{evolaa}
\big\langle A^i(x)  A^i(y)\big\rangle_{b\Lambda}&=&
\Big\langle\big\langle({\cal A}^i+\delta A^i)_x
({\cal A}^i+\delta A^i)_y\big\rangle_\rho\,\Big\rangle_{W_{\Lambda}}\nn
&\equiv & \big\langle{\cal A}^i(x){\cal A}^i(y)\big\rangle_{W_{b\Lambda}},\eeq
where the internal brackets $\langle\cdots\rangle_\rho$ in the
first line stand for the quantum average over semi-fast 
fluctuations (cf. eq.~(\ref{strip})) at fixed $\rho$, while
the external brackets $\langle\cdots\rangle_{W_{\Lambda}}$ denote
the classical average over $\rho$ with weight function $W_\Lambda[\rho]$.
By definition, the result of this double averaging (i.e., of the
classical plus quantum calculation) in the original effective theory at the
scale $\Lambda^+$ must be the same as the result of a purely
classical calculation in the new effective theory at the scale $b\Lambda^+$.
This is the content of the second line in eq.~(\ref{evolaa}).

The quantum corrections in the first line of eq.~(\ref{evolaa})
must be computed to lowest order in $\alpha_s\ln(1/b)$,
but to all orders in the background fields ${\cal A}^i$. 
This is essentially an one-loop calculation, but with the exact
background field propagator of the semi-fast gluons.
By matching its result with the classical calculation in the second 
line in eq.~(\ref{evolaa}), one can deduce\cite{PI}
the functional change $\Delta W \equiv W_{\tau+\Delta\tau} - W_\tau$
necessary to absorb these new correlations.
Since $\Delta W\propto \Delta \tau$,
the evolution is formulated as a (functional) RGE, which
is most conveniently written for $W_\tau[\alpha]\equiv
W_\tau[\rho=- \nabla^2_\perp \alpha]$, cf. eq.~(\ref{alpharho}).
It reads\cite{PI} \footnote{In Ref.\cite{W}, Weigert has shown
that Balitsky's equations\cite{B} can be summarized into
a functional equation equivalent to (\ref{RGEA}). More recently,
a simplified derivation of
eq.~(\ref{RGEA}) has been given by Mueller, from the dipole point of 
view\cite{Al01}.} :
\be\labe{RGEA}
{\del W_\tau[\alpha] \over {\del \tau}}\,=\,\int_{x_\perp,\,y_\perp}
{1 \over 2}\, {\delta \over {\delta
\alpha_\tau^a(x_\perp)}}\,\eta_{ab}(x_\perp,y_\perp)\, 
{\delta \over { \delta \alpha_\tau^b(y_\perp)}} W_\tau
[\alpha]\,\equiv\,-H W_\tau[\alpha],\ee
where $\eta_{ab}(x_\perp,y_\perp)$ it itself a
non-linear functional of the color field $\alpha_a(x)$, 
via the Wilson lines $V$ and $V^\dagger$ defined in eq.~(\ref{v}) :
\be\label{eta}
\eta^{ab}(x_\perp,y_\perp)
={1\over \pi}\int {d^2z_\perp\over (2\pi)^2}
\frac{(x^i-z^i)(y^i-z^i)}{(x_\perp-z_\perp)^2(y_\perp-z_\perp)^2 }
\Bigl\{1+ V^\dagger_x V_y-V^\dagger_x V_z - V^\dagger_z V_y\Bigr\}^{ab}.
\ee
The functional derivatives
in eq.~(\ref{RGEA}) are taken with respect to 
$\alpha^a_\tau(x_\perp)\equiv
\alpha^a(x^- = x^-_\tau,x_\perp)$, where
$x^-_\tau \equiv 1/\Lambda^+ \sim {\rm e}^{\tau}$.
This is so since the quantum corrections are located in the
strip $1/\Lambda^+ \simle x^- \simle 1/b\Lambda^+$, i.e.,
on top of the original field which has
support at $0\le x^-\simle1/\Lambda^+ $.
Thus, by integrating out the quantum modes in layers of $p^+$,
one constructs the classical field (or source) in layers of $x^-$,
with a one-to-one correspondence between $p^+$ and $x^-$
which reflects the uncertainty principle
$\Delta x^-\Delta p^+\sim 1$.

\section{General properties and consequences}
\label{sect:RGE}

Eq.~(\ref{RGEA}) has the structure of a diffusion equation: It
is a second-order (functional) differential equation whose r.h.s.
is a total derivative, as necessary to conserve the total
probability. It describes
quantum evolution as the diffusion (with ``time'' $\tau$)
of the probability density $W_\tau[\alpha]$
in the functional space spanned by $\alpha_a(x^-,x_\perp)$.

For comparison, consider the usual Fokker-Planck equation describing 
Brownian motion in flat space ($P(x,t)\equiv$ the probability density
to find the particle at point $x$ at time $t$
knowing that it was at $x=0$ at time 0) :
\be\label{FPBM}
{\del P(x,t)\over {\del t}}\,=\,D{\del^2\over \del x^i\del x^i}\,P(x,t)\,-\,
{\del\over \del x^i}\Bigl(F^i(x) P(x,t)\Bigr),\ee
where $D$ is the diffusion constant
and $F^i=-{\del V/ \del x^i}$ is an external force.
If $F^i=0$, the corresponding solution:
\be\label{DIFF} P(x,t)\,=\,{1\over (4\pi Dt)^{3/2}}\,\,
{\rm exp}\Big\{-\frac{x^2}{4Dt}\Big\}\ee
goes smoothly to zero at any $x$ when $t\to \infty$ (runaway solution).
But for $V\ne 0$, there exists a non-trivial stationary solution 
$P_0(x)\sim {\rm exp}[-V(x)/D]$, i.e., an 
equilibrium distribution that is asymptotically reached by the system.

There is manifestly no force term in the RGE (\ref{RGEA}). This
property 
relies on subtle compensations between real and virtual corrections
\cite{W,PI}, and entails a purely diffusive behaviour, as
in eq.~(\ref{DIFF}). Thus, there is no fixed point,
or stationary distribution, for the flow described by eq.~(\ref{RGEA}).

From the functional RGE (\ref{RGEA}), one can derive ordinary
evolution equations for all the observables which 
can be computed as an average over $\alpha\,$.
Particularly interesting quantities are the gauge-invariant
products of Wilson lines $\langle
{\rm tr}(V^\dagger_x V_y){\rm tr}(V^\dagger_z V_u)\cdots \rangle_\tau$,
for which eq.~(\ref{RGEA}) predicts the
same evolution equations 
as obtained by Balitsky\cite{B} via operator product expansion.
In particular, the 2-point function
$\langle {\rm tr}(V^\dagger_x V_y)\rangle_\tau$ satisfies
\beq\labe{evolV}
{\del \over {\del \tau}}\langle {\rm tr}(V^\dagger_x V_y)
\rangle_\tau&=&-{\alpha_s\over 2 \pi^2}\int d^2z_\perp
\frac{(x_\perp-y_\perp)^2}{(x_\perp-z_\perp)^2(y_\perp-z_\perp)^2 }\nn
&{}&\qquad\quad\times
\left\langle N_c {\rm tr}(V^\dagger_x V_y)
- {\rm tr}(V^\dagger_x V_z){\rm tr}(V^\dagger_z V_y)\right\rangle_\tau\eeq
which in the large $N_c$ limit becomes a closed equation, due to 
Kovchegov\cite{K}, that can be recognized as a
non-linear generalization of BFKL.
Its solution determines the 
cross section $\sigma_{dipole}(\tau,r_\perp)$
for the scattering of a ``color dipole'' of size
$r_\perp=x_{\perp}-y_{\perp}$ off the hadron \cite{AM0,K}.
Recent progress\cite{Levin01} with this equation 
exhibits color transparency ($\sigma_{dipole}\propto
r_\perp^2 Q_s^2(\tau)$)  at small distances $r_\perp \ll 1/Q_s(\tau)$, 
and saturation  ($\sigma_{dipole}\approx $ const) at large
distances $r_\perp \gg 1/Q_s(\tau)$, in qualitative agreement
with a phenomenological model by
Golec-Biernat and W\"usthoff \cite{GBW} which
successfully describes the HERA data.
In the above estimates, $Q_s(\tau)$ is the saturation scale to
be discussed in the next section.

\section{Gluon Saturation and Unitarity}

Except in the large $N_c$ limit, Balitsky's equations do not
close individually, but rather form an infinite hierarchy of coupled
equations. It is then more convenient to study directly the functional RGE
(\ref{RGEA}). As shown in Ref.\cite{SAT}, approximate
solutions to this equation can be obtained
in limiting kinematical regimes, by combining
mean field approximations (MFA) and kinematical simplifications.
Moreover, an exact solution, in the form of a path-integral
in 2+1 dimensions, can be also written down\cite{BIW}, which may be
used in lattice calculations. (See also the second Ref.\cite{B} for
a different path-integral formulation.)
Here, I shall briefly describe the  analytic solutions, 
with emphasis on the phenomenon of gluon saturation\cite{SAT}.

The MFA consists in replacing $\eta_{ab}(x_\perp,y_\perp)$ 
in the kernel of the RGE (\ref{RGEA}) by its expectation value
$\langle\eta_{ab}(x_\perp,y_\perp)\rangle_\tau\equiv
\delta^{ab}\gamma_\tau(x_\perp,y_\perp)$. Then, the 
equation can be solved exactly, and the solution $W_\tau$
--- a  Gaussian in $\alpha$ (or $\rho$) --- can then be used
to self-consistently determine the approximate kernel 
$\gamma_\tau(x_\perp,y_\perp)$.

The kinematical approximations rely on the fact that there is an
intrinsic transverse scale in the problem, the saturation scale
$Q_s(\tau)$, which is the inverse correlation length for Wilson lines:
\be\label{NT0}
\langle V^\dagger_x V_y\rangle_\tau\,\approx
\left\{ \begin{array} {c@{\quad\rm for\quad}l}
1, & |x_{\perp}-y_{\perp}|\ll 1/Q_s(\tau) \\
0, & |x_{\perp}-y_{\perp}|\gg 1/Q_s(\tau)
\end{array}
\right.
\ee
To get an estimate for $Q_s(\tau)$, one starts the analysis at
sufficiently short distances $|x_{\perp}-y_{\perp}|\ll 1/Q_s(\tau)$,
where the fields are weak and their dynamics is linear
($V^\dagger_x\approx 1+ig\alpha(x_{\perp})$), and study the onset
of non-linearities with increasing $|x_{\perp}-y_{\perp}|$.
In this regime, one obtains (cf. eq.~(\ref{GCL})):
\be\label{Vhigh}
\langle V^\dagger_x V_y\rangle_\tau\,\simeq\,
{\rm exp}\Big\{-{\bar\alpha_s\over 2}{r_\perp^2\over R^2}\,
xG\big(x,1/r_\perp^2\big)\Big\},\ee
($\bar\alpha_s=\alpha_s N/\pi$,
$r_\perp=x_{\perp}-y_{\perp}$, and $R$ the hadron radius)
with $xG(x,Q^2)$ satisfying the standard evolution equation in the double
log approximation:
\be\label{DLA}
{\partial^2\over \partial \tau\,\partial\log Q^2
}\,xG(x,Q^2)\,
=\,\bar\alpha_s xG(x,Q^2).\ee
(Here, this equation arises via the  self-consistency requirement.)
Non-linear effects, which are negligible at sufficiently small $r_\perp$,
become important for $r_\perp\sim 1/Q_s(\tau)$ where the exponent
in eq.~(\ref{Vhigh}) becomes of order one. Thus,
\be\label{QSHM}
R^2Q^2_s(\tau)\,\simeq\,{\bar\alpha_s}\,
xG\big(x,Q^2_s(\tau)\big),\ee
which together with eq.~(\ref{DLA}) 
implies that $Q_s(\tau)$ increases exponentially:
\be\label{Qstau}
Q^2_s(\tau)\,=\,{Q^2_s(\tau_0)\,\rm e}^{\,c\bar\alpha_s(\tau-\tau_0)}\,
\qquad (c=4\,\,{\rm in\,\,DLA}).
\ee

The weight function at high-momenta $k_\perp
\gg Q_s(\tau)$ is the Gaussian:
\be \label{Whigh}
W^{{\rm high}-k_\perp}_\tau[\rho]\,\simeq\,
{\rm exp}\Big\{-{1\over 2}\int_{Q_s(\tau)} {d^2k_\perp\over 
(2\pi)^3}\,\frac{\rho^a(k_\perp)\rho^a(-k_\perp)}{\mu_\tau(k_\perp)}
\Big\},\ee
where $\mu_\tau(k_\perp)\propto
 (\partial/\partial\log k_\perp^2) xG(x,k_\perp^2)$ is only slowly varying
with $k_\perp$. Eq.~(\ref{Whigh}) is the McLerran-Venugopalan 
model of independent colour charges\cite{MV94,Larry01},
amended by the standard, linear, quantum evolution, cf. eq.~(\ref{DLA}).

The new physics, intrinsically non-linear, shows up at large distances
($r_{\perp}\gg 1/Q_s(\tau)$), or low momenta
 ($k_\perp \ll Q_s(\tau)$, with $k_\perp \gg \Lambda_{QCD}$ though), 
where color fields are strong, $\alpha^a\sim 1/g$, and Wilson
lines are decorrelated, $V^\dagger_x V_y\approx 0$. Then one can neglect
the bilinears involving Wilson lines in eq.~(\ref{eta}),
so that the corresponding weight function is very simple:
\be\label{Wlow}
W_\tau^{{\rm low}-k_\perp}[\rho]\,\simeq\,
{\rm exp}\Big\{-{1\over 2}\int^{Q_s(\tau)} {d^2k_\perp\over 
(2\pi)^3}\,\frac{\rho^a(k_\perp)\rho^a(-k_\perp)}{
\varepsilon_\tau(k_\perp)k_\perp^2}
\Big\}.\ee
This describes a 2-dimensional Coulomb gas, with ``dielectric
constant''
\be
\varepsilon_\tau(k_\perp)\,\equiv\,\tau\,-\,\bar\tau(k_\perp)
\,=\,{1\over c\bar\alpha_s}\,
\ln{Q_s^2(\tau)\over k_\perp^2}\,,\ee
which is the rapidity window for quantum evolution in the
saturation regime. Indeed, for given $k_\perp \ll Q_s(\tau)$,
only  those partons are saturated whose rapidities y are
large enough for 
$Q_s({\rm y}) > k_\perp$. This requires $\bar\tau(k_\perp) <
{\rm y} <\tau$, with $\bar\tau(k_\perp)$ the rapidity where
$Q_s\big(\bar\tau(k_\perp)\big)=k_\perp$.

Saturation is an immediate consequence 
of eq.~(\ref{Wlow}), which implies the
following gluon density per unit transverse phase space
at $k_\perp < Q_s(\tau)$ :
\be\label{SAT}
\frac{d^2(x G )}{d^2k_\perp\,d^2b_\perp}={N^2-1\over 4 \pi^4}\,
\bigl( \tau-\bar\tau(k_\perp)\bigr)=
{N^2-1\over 4 \pi^4 c}\,{1\over \bar\alpha_s}\,
\ln{Q_s^2(\tau)\over k_\perp^2}\,.\ee
At fixed $\tau$,
the gluon density is almost constant\footnote{Eq.~(\ref{SAT}) shows 
only a mild,
logarithmic, dependence upon $k_\perp$, to be contrasted with the
$1/k_\perp^2$ behaviour at large $k_\perp$, due to bremstrahlung.}
for $k_\perp < Q_s(\tau)$, and of order
${\cal O}(1/\alpha_s)$. This is the Color Glass Condensate.
When $\tau\sim\ln s$ increases, the gluon density at
$k_\perp < Q_s(\tau)$ increases only linearly with $\tau$ (i.e.,
logarithmically with the energy $s$), but the saturation
scale  $Q_s(\tau)$ itself increases as a power of $s$. That is, with 
increasing energy, the new partons are predominantly produced at large
transverse momenta $\simge  Q_s(\tau)$, so they cannot be
discriminated by an external probe with resolution $Q^2<Q_s^2(\tau)$.
Thus, although the total number of gluons 
keeps growing when the energy increases, there is no contradiction
with unitarity since, beyond some energy, the partons produced
by quantum evolution are too tiny to
contribute to the cross section at fixed $Q^2$.
Saturation is a natural mechanism to restore unitarity.

\end{document}